\documentclass[a4paper,fleqn]{cas-dc}

\usepackage[numbers]{natbib}

\usepackage{graphicx}
\usepackage{amsmath}
\usepackage{algorithm}
\usepackage{algorithmic}
\usepackage{xurl}
\usepackage{todonotes}
\graphicspath{ {./figures/} }
\usepackage{subcaption}
\usepackage{booktabs}
\usepackage{siunitx}
\usepackage{enumitem}
\usepackage[dvipsnames]{xcolor}

\usepackage{soul}

\newcommand{\blindtext}[1]{\textbf{$<<$Text hidden for blind review$>>$}}
\newcommand{\review}[1]{\textcolor{black}{#1}}
\newcommand{\rebuttal}[1]{\textcolor{blue}{#1}}

\newcommand{\green}[1]{\textcolor{ForestGreen}{#1}}
\newcommand{\red}[1]{\textcolor{red}{#1}}

\usepackage[percent]{overpic}
\usepackage[super]{nth}
\usepackage{pdfpages}
\usepackage{multirow}

\DeclareUnicodeCharacter{2009}{\,}

\begin{document}

\let\WriteBookmarks\relax
\def\floatpagepagefraction{1}
\def\textpagefraction{.001}

\shorttitle{MPI Malleability Validation under HPC Conditions}    

\shortauthors{S. Iserte et al.}  

\title [mode = title]{MPI Malleability Validation under Replayed Real-World HPC Conditions}

\author[1]{Sergio Iserte}[orcid=0000-0003-3654-7924]
\author[2]{Maël Madon}[orcid=0000-0001-9476-4682]
\author[2]{Georges Da~Costa}[orcid=0000-0002-3365-7709]
\author[2]{Jean-Marc Pierson}[orcid=0000-0001-8948-0474]
\author[1]{Antonio J. Peña}[orcid=0000-0002-3575-4617]
\ead{antonio.pena@bsc.es}
\cormark[1]
\cortext[1]{Corresponding author}

\affiliation[1]{organization={Barcelona Supercomputing Center (BSC)},
            city={Barcelona},
            country={Spain}}
            
\affiliation[2]{organization={University of Toulouse, CNRS, Toulouse INP, IRIT},
            city={Toulouse},
            country={France}}




\begin{abstract}
Dynamic Resource Management (DRM) techniques can be leveraged to maximize throughput and resource utilization in computational clusters.
Although DRM has been extensively studied through analytical workloads and simulations, skepticism persists among end administrators and users regarding their feasibility under real-world conditions.
To address this problem, we propose a novel methodology for validating DRM techniques, such as malleability, in realistic scenarios that reproduce actual cluster conditions of jobs and users by replaying workload logs on a High-performance Computing (HPC) infrastructure.
Our methodology is capable of adapting the workload to the target cluster.
We evaluate our methodology in a malleability-enabled 125-node partition of the Marenostrum~5 supercomputer. Our results validate the proposed method and assess the benefits of MPI malleability on a novel use case of a pioneer user of malleability (our ``PhD Student''): parallel-efficiency-aware malleability reduced a malleable workload time by 27\% without delaying the baseline workload, although introducing queueing delays for individual jobs, but maintaining the resource utilization rate.
\end{abstract}



\maketitle

\section{Introduction}\label{intro}
High-performance computing (HPC) facilities are critical for advancing scientific research, engineering, and data analysis across various domains, from genomic research~\cite{martinez_dynamic_2013, zhong_gpu_2025} to multi-physics simulations~\cite{hess_gromacs_2008, vazquez_alya_2016, caviedes-voullieme_serghei_2023} through artificial intelligence~\cite{martinez-cuenca_use_2023, rosciszewski_optimizing_2023, godoy_large_2024}. 
These computational clusters rely on efficiently utilizing resources to maximize their productivity. Dynamic resource management (DRM) techniques have emerged as key strategies to improve the utilization of these systems~\cite{bungartz_invasive_2013, garcia_hints_2014, lopez_openmp_2021}. These techniques enable the flexible allocation and reallocation of computational resources and reconfiguring jobs accordingly to adapt to the dynamic nature of HPC workloads.
Among others, malleability based on the Message-Passing Interface (MPI) is one of the most extended DRM approaches to process layout reshape and data redistribution
~\cite{10.1145/3555819.3555856, iserte_agut_high-throughput_2018, martin_flex-mpi_2013, bhattarai_dynamic_2024}.

Despite their potential, DRM techniques face several challenges that hinder their adoption in production environments~\cite{iserte_resource_2025}.
Particularly, MPI malleability requires 1) a malleability-ready resource management system (RMS), which many existing facilities do not possess, and 2) a malleability framework to develop malleable applications compatible with the RMS and the available distributed parallel runtime systems in the cluster.
These kinds of limitations increase the complexity of evaluating DRM techniques with realistic workloads and foster skepticism among end users and system administrators regarding DRM's practical benefits and feasibility.
\review{
    In this regard, previous studies evaluating DRM techniques have relied on simplistic simulations, biased benchmarks, or synthetic traces, often failing to accurately represent the conditions of real-world systems. For example, some works limit their evaluations to small-scale testbeds~\cite{sudarsan_reshape_2007,sarood_maximizing_2014, prabhakaran_batch_2015}, which struggle to provide a nice picture of the queuing dynamics or contention effects of production supercomputers. Others employ synthetic workloads or with uniform job sizes or idealized submission patterns~\cite{sudarsan_reshape_2007, iserte_dmr_2018}, overlooking the variability and burstiness of user behavior. Still others focus only on single-application benchmarks or simulators~\cite{sarood_maximizing_2014, sudarsan_dynamic_2009, iserte_study_2020, iserte_dynamic_2019}, which, while useful to demonstrate mechanisms, do not provide insight into system-wide interactions with the RMS. These shortcomings have limited the impact of prior results, making it difficult for the community to assess the practical benefits of DRM techniques. A more detailed analysis of these studies is provided in Section~\ref{sec:related}.
}

To address the challenges of evaluating DRM techniques, we propose a novel methodology that leverages supercomputer logs to generate realistic workloads based on actual user submission behaviors, including temporal patterns across hours and days of the week. Our approach introduces a mechanism called the \textit{User-Based Submitter}, designed to scale across different computing clusters. Rather than relying on synthetic simulations, this method replays real user interactions from production supercomputers, enabling a more accurate and comprehensive evaluation of DRM techniques.

We further incorporate malleable jobs into these replayed workloads to assess the impact of DRM on job execution and system performance. This allows us to observe the effects of job malleability and resource reallocation in a controlled yet realistic environment. Notably, the study includes the perspective of a pioneering user---submitting malleable jobs in a 125-node partition---demonstrating the potential of our methodology to validate previous findings and uncover new insights in real HPC settings. It is worth noting that the cluster size used in our experiments ranks among the largest evaluated in state-of-the-art malleability studies (see Section~\ref{sec:related}).

In summary, this paper presents 
1) \textbf{a methodology to replay logs in any computational cluster keeping the logic behind user submissions}. 
Instead of a mathematical model, this methodology uses the techniques of user sampling from a recorded log and replay with feedback to reproduce the workload in a target system (see Section~\ref{sec:methodology}).
This paper particularly 
2) \textbf{designs and analyzes a novel case study of the adoption of malleability in a production system}. For this purpose, the authors exemplify this event with a new user (``the student'') who is the pioneer in submitting malleable jobs to the cluster. Among all the possible scenarios, the authors have decided to develop a case that paves the way to more complex malleable workloads (see Section~\ref{sec:experiments}).
Similarly to hardware simulators that have to be validated in actual hardware to be published, this paper presents, for the first time in the literature, 
3) \textbf{the validation of MPI malleability using a real workload in a malleability-enabled supercomputer}, without simulations or synthetic benchmarks (see Section~\ref{sec:results}). 

The paper ends with a discussion of the applicability of the presented technique and the meaningful results obtained in the evaluation of the validation methodology (see Section~\ref{sec:discussion}). It concludes with a summary of our contributions and findings (see Section~\ref{sec:conclusion}).

\rebuttal{
    At the end of the day, we aim to bridge the gap between simulation-based studies and practical deployment in HPC infrastructures.}

\section{Related Work}\label{sec:related}
This section reviews the previous efforts done in the fields of workload replay and dynamic resource management.

\subsection{Workload Replay}
A classical approach to evaluate resource management policies consists of replaying a historical workload from its log.
However, modifying the target infrastructure or scheduling policies impacts the system's performance (e.g., response time, computing speed, resource availability),
which has a crucial impact on the submission behavior of its users~\cite{zakay_preserving_2014, schlagkamp_influence_2017}.
As a result, many previous works in the field have generalization issues, since they do not account for these behaviors.
A solution is to perform closed-loop simulations~\cite{schroeder_open_2006},
where the submitted workload adapts to the simulated performance of the system.
To do so, the authors in~\cite{zakay_identifying_2013} extract relevant submission patterns from historical workloads to replay them in simulations.
More recently, the authors in~\cite{madon_replay_2024} provide an in-depth study of this technique, named ``replay with feedback''.
In~\cite{feitelson_resampling_2021}, the authors go one step further and suggest using \emph{resampling} with feedback for performance evaluation where the historical workload is scaled to the target infrastructure through user sampling.

\subsection{Dynamic Resource Management}
The MPI paradigm is the {\it de facto} standard for implementing distributed parallel applications designed to run on computational clusters. MPI provides efficient and straightforward mechanisms for communication among different processes, known as ranks, each with its own memory address space.
In a traditional MPI job, a fixed number of ranks are initiated at the start and remain active until the completion of the job. This static allocation of resources may lead to inefficiencies, especially in dynamic and heterogeneous computing environments.

MPI malleability addresses these inefficiencies through DRM. This technique allows the number of MPI ranks to be modified during the execution of a program, enabling the application to resize on the fly~\cite{feitelson_packing_1996}. Malleable MPI applications may adapt to changing computational resources, improving overall system utilization and application performance.
Although MPI malleability could be seen as a variant of checkpoint/restart, most modern solutions implement on-memory data redistribution, reducing overheads from accessing storage media.

Closely related to malleability is the concept of moldability. While malleability involves adjusting the number of MPI ranks during the execution of a job, moldability determines the number of ranks at the time of job submission before its initialization~\cite{lublin_workload_2003}. 
In this regard, jobs specify a range of sizes, allowing the job scheduler to choose the most suitable configuration based on current system availability and load.
This approach provides an extra degree of flexibility by enabling better resource allocation at job scheduling time, which can lead to improved system efficiency and potentially shorter wait times for job initiation. However, unlike malleable applications, moldable applications do not adjust their resource usage dynamically during runtime.

Surveys in ~\cite{aliaga_survey_2022, tarraf_malleability_2024} extensively review the state--of--the--art of malleability in HPC systems. Instead, we focus on \textbf{how malleability frameworks and actual RMS have been evaluated with workloads} to demonstrate their usefulness of DRM in high-performance clusters.

ReSHAPE is a coupled solution for adaptive workloads, including its specific reconfiguration libraries, scheduler, and runtime system.
This strong integration forces ReSHAPE users to develop applications that are compatible with this system.
In~\cite{sudarsan_reshape_2007}, ReSHAPE is evaluated in a 50-node cluster with a 5-job workload composed of the benchmarks LU, MM, Master-worker, Jacobi, and FFT, submitted simultaneously.
A subsequent publication added to the workload job instances of a malleable version of LAMMPS~\cite{sudarsan_dynamic_2009}.

PARM (acronym of Power-Aware Resource Manager) relies on over-provisioning, power capping, and job malleability, based on CHARM++ and Slurm~\cite{sarood_maximizing_2014}.
PARM is evaluated in a 38-node cluster with a 5-job workload comprising the benchmarks and applications Wave2D, Jacobi2D, LeanMD, Lulesh, and AMR.
Larger experiments relied on the Slurm simulator with a log of 68,936 jobs submitted to a 40,960-node cluster. The authors assume that all the jobs in the workload are malleable and assign them random scalability from a set of samples.

The solution developed by Prabhakaran et al. combines AMPI (based on CHARM++) with the RMS Torque/Maui to tackle malleable jobs~\cite{prabhakaran_batch_2015}. 
The authors evaluate in a 15-node cluster a workload based on a modified version of the ESP benchmark to contain various percentages of rigid, malleable, and evolving jobs composed of 230 instances.
While the rigid jobs followed the benchmark directives, the flexible jobs were a malleable version of LeanMD and an evolving implementation of the synthetic Quadflow.

The Dynamic Management of Resources Library (DMRlib) (detailed in Section~\ref{subsec:dmr}) was evaluated in a 129-node cluster with four benchmarks and applications (CG, Jacobi, N-Body, HPG-Aligner~\cite{iserte_dynamic_2019}) instantiated in 2,000-job workloads with various percentages of malleable jobs and turning on and off malleability in the applications of the workloads~\cite{iserte_dmrlib_2020}.
The authors leveraged the Feitelson synthetic workload generator~\cite{feitelson_towards_1996} to evaluate job scheduling algorithms in HPC.
This generator provides a way to simulate job arrival patterns, job sizes, and resource demands based on empirical data from real-world systems.

While prior research has explored the potential of DRM through benchmarks and simulations, these studies often rely on synthetic workloads with arbitrarily defined job sizes and submission patterns—such as steady or bursty arrivals—that fail to capture the complexity and variability of real-world HPC usage. These simplified models limit the generalizability and applicability of the findings to production systems.

\rebuttal{
    In contrast, our work introduces a \textbf{methodology that replays actual user} behavior extracted from supercomputer logs, preserving the nuanced temporal and structural characteristics of real workloads. This approach \textbf{enables the validation of past and future synthetic setups} and their results, overcoming the limitations of synthetic modeling where user behavior is too complex to be accurately captured by mathematical functions. 
}

\section{Methodology}\label{sec:methodology}

In order to study DRM under real-world conditions,
our method consists of reproducing the nominal activity of a cluster computing infrastructure.
We use a workload submitter based on the logic of users (Section~\ref{subsec:submitter}).
The nominal activity is represented by submitting non-malleable jobs.
These jobs, later called ``baseline workload'', are reproduced from historical workload logs and sized to our testbed through user sampling (Section~\ref{subsec:sampling}).
To study the effects of DRM on top of the baseline workload, the submitter includes an additional user leveraging moldability and malleability in her jobs, thanks to a malleable runtime (Section \ref{subsec:dmr}).

\subsection{User-Based Workload Submitter}
\label{subsec:submitter}

The core of the experiments presented in this paper relies on a piece of software running throughout the whole experiment duration: the User-Based Submitter (UBS)\footnote{Code available at \url{https://gitlab.bsc.es/siserte/ubs}}.
UBS reproduces HPC users issuing job submissions to the Slurm resource manager at certain timestamps.
There are two types of users as described below:
\begin{itemize}[wide]
    \item \textit{Traditional users} submit jobs whose execution time, number of parallel resources, and submission time are given in an input file, based on a historical workload.
    These are used to replay the baseline workload.
    Consequently, the submitted jobs replicate the traditional rigid workload, allocating the requested resources during the desired time.

    \item \textit{Generative users} submit malleable jobs, using feedback on the status of previous submissions to decide on the following job submissions (see \cite{madon_replay_2024}).
    More precisely, a generative user takes as input a triplet $(t_0, \Delta t, N)$
    with $t_0$ the timestamp of their first submission,
    $\Delta t$ the think time between the end of the previous job and the submission of the new job and $N$ the total number of submissions from this user.
\end{itemize}

In the UBS, traditional and generative users are managed in two dedicated threads. 
The thread managing the traditional users sleeps until the next submission time, then calls a shell script issuing the job submission, passing it all relevant parameters about the job (i.e.: type, duration, or number of resources).
The thread of the generative users uses a socket to be informed of the completion of previous jobs. 
When a job owned by a generative user terminates, a new thread is started, sleeping for $\Delta t$ seconds before calling the submission script.

\subsection{User Sampling}
\label{subsec:sampling}

\renewcommand{\algorithmicrequire}{\textbf{Input:}}
\renewcommand{\algorithmicensure}{\textbf{Output:}}

\begin{figure*}[tbp]
  \centering
  \begin{minipage}{.6\linewidth}
    \begin{algorithm}[H]
    \caption{User sampling from workload}\label{alg:sampling}
    \begin{algorithmic}[1]
        \REQUIRE $\bar{m} =$ the target median/average load per day
        \REQUIRE $M =$ the number of nodes in the target platform

        \STATE $U \leftarrow$ list of users in the workload
        \STATE $pool \leftarrow \{\}$, $m \leftarrow 0$
        \WHILE{$m \notin [.95\bar{m}, 1.05\bar{m}]$}
            \STATE $u \leftarrow$ draw a random user from $U$, without replacement 
            \IF{$max($nodes required by $u) < M$}
                \STATE $pool$.add($u$)
                \STATE $m' \leftarrow$ load per day from users in $pool$
                \IF{$m' > 1.05\bar{m}$}
                    \STATE $pool$.remove($u$)
                \ELSE
                    \STATE $m \leftarrow m'$
                \ENDIF
            \ENDIF
        \ENDWHILE
        \IF{$m \notin [.95\bar{m}, 1.05\bar{m}]$}
            \STATE print("SAMPLING FAILED!")
        \ENDIF

        \ENSURE $pool$
    \end{algorithmic}

    \end{algorithm}
  \end{minipage}
\end{figure*}

To create a realistic baseline workload of the actual activity in a cluster platform,
we leverage historical workload logs.
The challenge is to adapt this workload to the size of our testbed
\textit{without losing the logic of submission of the original users}.
Thus, instead of randomly sampling jobs from the original log,
we rather perform a sampling of \textit{users}, 
proceeding as described below (see Algorithm~\ref{alg:sampling}). 

First, we choose a target level of activity for the platform,
in terms of average or median load submitted per day.
For example, if the testbed has 100 nodes and we target 85\% activity,
we want the baseline workload to have approximately $85\times 24$ node-hours per day.
To reach the level, we proceed iteratively by randomly adding users to a user pool.
At each step, we calculate the load submitted daily by the users in the pool
by summing the product of execution time and number of nodes for all jobs submitted by these users in the original log.
If the activity reaches the desired level (with a margin of plus or minus 5\%), the process is stopped; 
otherwise, we add a new user.
The last added user may overshoot the target by more than 5\%;
in this case, that user is removed from the pool before adding another.

\subsection{Dynamic Resource Management}\label{subsec:dmr}

\begin{figure*}
\centering
    \includegraphics[clip,width=0.8\linewidth,trim={1.2cm 1.2cm 1.2cm 0.25cm}]{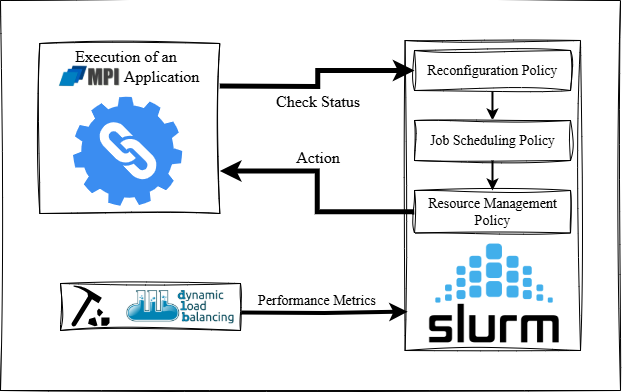}
    \caption{DMRlib Application--MPI--Slurm communication.}
    \label{fig:dmr-slurm}
\end{figure*}

This research enables DRM by leveraging moldability and malleability thanks to the Dynamic Management of Resources Library (DMRlib)~\cite{iserte_dmrlib_2020}.
DMRlib is a high-level API that facilitates the adoption of malleability in HPC codes. 
DMRlib implements a communication layer between the parallel distributed runtime (PDR) and the RMS, driving the management of processes and resources transparently to the user while providing flexibility to increase productivity and resource utilization of HPC facilities.

Figure~\ref{fig:dmr-slurm} depicts the operation of dynamic resource management in DMRlib, which is described as follows:

\begin{itemize}
\item During execution, jobs periodically expose their readiness for reconfiguration to the RMS.
\item This communication occurs at a synchronization point specified in the code, where the reconfiguration process may commence. In iterative applications, the end of an iteration often serves as an ideal synchronization point.
\item The RMS, with its cluster-wise information and the performance metrics provided by a monitor, determines reconfiguration actions following its defined policies, and informs the PDR.
\item If this results in a change to the job size, the RMS reallocates resources and returns the new number of processes, which may involve either expanding or shrinking the job.
\item Finally, the PDR redistributes the data among the processes according to the application's guidelines, allowing jobs to continue execution with the new process layout from the point where the reconfiguration was triggered.
\end{itemize}

\review{
DMRlib supports multiple Process Dynamic Runtimes (PDRs), including Nanos++~\cite{iserte_dmr_2018} and several MPI implementations~\cite{iserte_malleable_2025}, and can seamlessly interface with different malleability backends such as Proteo~\cite{martin-alvarez_proteo_2024} and DPP~\cite{huber_bridging_2025}.
}

In this work, DMRlib has been configured with MPICH\footnote{\url{https://www.mpich.org}}, as the PDR, and Slurm\footnote{\url{https://slurm.schedmd.com}} as the RMS~\cite{iserte_towards_2025}.
DMRlib integrates a monitoring module to track performance metrics during job execution and react accordingly by requesting/suggesting expansion or shrinking actions to the RMS.
Metrics are gathered by TALP~\cite{lopez_talp_2021}, which collects ``POP metrics''\footnote{\url{https://pop-coe.eu/node/69}}. This set of metrics is organized hierarchically and is multiplicative; that is, the value of the parent metric is equal to the product of the child metrics.
Among others, Parallel Efficiency (PE) reveals the inefficiency in splitting computation over processes and then communicating data among processes.
PE is a compound metric whose components reflect two critical factors to attain good parallel performance by ensuring even distribution of computational work across processes (load balance) and minimizing time communicating data among processes (communication efficiency).


\section{Experimental Setup}\label{sec:experiments}
We implement the method described above to evaluate the impact on the job submissions of a user enabling DRM in a computing cluster.
For the purpose of this narrative, we portray this user as a PhD student working under a tight deadline, submitting malleable instances of a scientific application.
This section describes the testbed, workload, and specific parameters of the experimental campaign.

\subsection{Cluster Computing Infrastructure}
\label{subsec:testbed}
The evaluation is performed in the general-purpose partition of the Marenostrum 5 (MN5) supercomputer, a pre-exascale machine integrated into the EuroHPC-JU European supercomputing infrastructure\footnote{\url{https://www.bsc.es/supportkc/docs/MareNostrum5/overview}}.
The nodes of this partition are based on Intel Sapphire Rapids (4\textsuperscript{th} Generation Intel Xeon Scalable Processors), each equipped with two Intel Xeon Platinum 8480+ with 56 cores running at 2~GHz of base frequency, for a total of 112 cores and 256~GB of DDR5 memory.
The nodes are interconnected through a 100~Gbit/s ConnectX-7 NDR200 InfiniBand network.

Regarding the software stack, the executions rely on the GCC-10.4 compiler, and the MPICH-3.2 MPI implementation.
\review{
We are aware of the current incompatibility between UCX and \texttt{MPI\_Comm\_spawn} in MPICH. For this reason, we opted to use OFI as the transport layer protocol instead.
At the same time, a new version of DMR (DMRv2) is under development, incorporating support for Open MPI and UCX. Preliminary results of this work have been published in~\cite{iserte_malleable_2025}.
}

The RMS is a customized version of Slurm to support malleability included in the dynamic resource management framework DMRlib. 
\review{The malleability-enabled Slurm provided in DMRlib is based on Slurm-17.02.0-0pre1 and extends its functionality through a resource selection plugin that supports malleability. The plugin builds upon the \texttt{select/linear} resource selection policy and remains fully compatible with Slurm’s native support for moldability. An extensive evaluation of this plugin, considering workloads with varying proportions of moldable and malleable jobs, is presented in~\cite{iserte_dmrlib_2020}.
That work concludes that moldability is not enough to reap the benefits of dynamic resource management, and malleability is required to achieve significant improvements in resource utilization and job throughput.
\\
In the current work, we have extended that plugin with performance metrics from TALP to drive malleability decisions as explained in Section~\ref{subsubsec:exps}.
For this reason, every time a malleable job triggers a reconfiguration, Slurm will determine the action that the runtime must perform, which may be \textit{expand}, \textit{shrink}, or \textit{none}. 
Notice that it is out of the scope of this work evaluate scheduling or priority policies, thus, Slurm is configured with \review{the default parameters of the} \texttt{sched/backfill} and \texttt{priority/multifactor} policies.
We have also enabled the Slurm job epilog mechanism to communicate job completions to the job submitter UBS described in Section~\ref{subsec:submitter}.
\\
The experiments are scoped within 125 nodes for up to 48 hours, which are the limits of the default QoS of the BSC researchers in MN5.
Slurm is deployed using a node running its controller daemon, while the remaining 124 nodes act as actual compute nodes.
}

\subsection{Traditional Users: The Baseline Workload}

\begin{figure*}[tbp]
    \centering
    \begin{subfigure}[t]{\linewidth}
        \centering
        \includegraphics[clip,width=.95\linewidth,trim={0.1cm 10.5cm 0.1cm 0.81cm}]{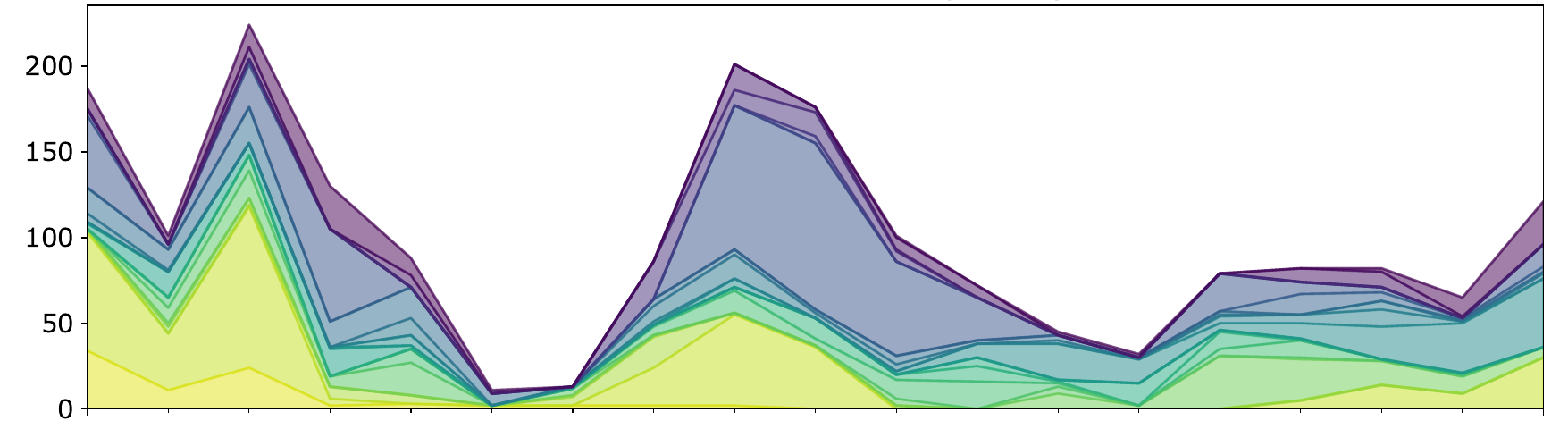}
        \caption{Number of submissions per day.}
        \label{fig:wl_distrib_sub_top}
    \end{subfigure}
    \vspace{1em} 
    \begin{subfigure}[t]{\linewidth}
        \centering
        \includegraphics[clip,width=.95\linewidth,trim={0.1cm .9cm 0.1cm 9.91cm}]{figures/submissions.pdf}
        \caption{Node-hours submitted per day.}
        \label{fig:wl_distrib_sub_bottom}
    \end{subfigure}
    \caption{Distribution of job submissions and platform capacity over the days of July 2017 (x-axis). Each color represents a different user. The horizontal line in the bottom graph is the maximum capacity of the platform ($M=124\times24=2,976$ node-hours per day).}
    \label{fig:wl_distrib_sub}
\end{figure*}

\begin{figure*}[tbp]
\centering
    \includegraphics[clip,width=.95\linewidth,trim={0.1cm 0.1cm 0.1cm 0.1cm}]{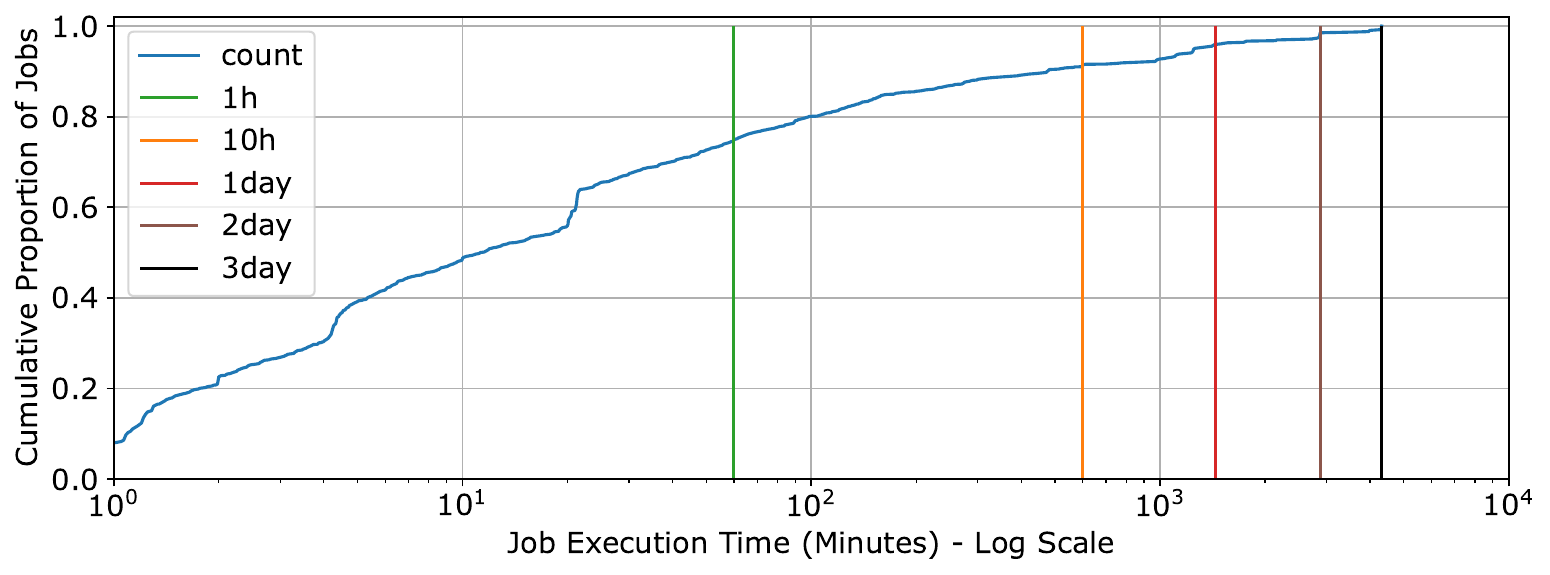}
    \caption{Distribution of job execution time in the baseline workload.}
    \label{fig:wl_distrib_cum}
\end{figure*}

\begin{figure*}[tbp]
\centering
    \begin{subfigure}[t]{\linewidth}
        \centering
        \includegraphics[clip,width=.95\linewidth,trim={0.1cm 10.5cm 0.1cm 0.8cm}]{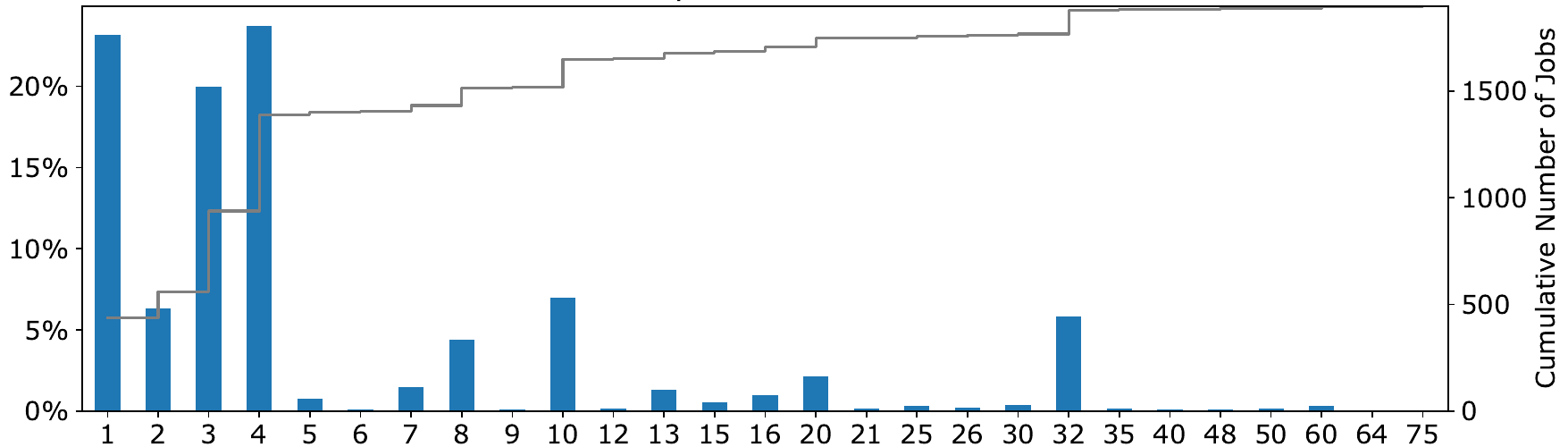}
        \caption{Proportion of jobs.}
        \label{fig:wl_distrib_runtime_size_top}
    \end{subfigure}

    \vspace{1em} 

    \begin{subfigure}[t]{\linewidth}
        \centering
        \includegraphics[clip,width=.95\linewidth,trim={0.1cm 0.1cm 0.1cm 10.5cm}]{figures/distrib_job_size.pdf}
        \caption{Proportion of node-hours.}
        \label{fig:wl_distrib_runtime_size_bottom}
    \end{subfigure}

    \caption{Job sizes in the baseline workload. The distribution is shown by the number of jobs (top) and number of node-hours (bottom), and the cumulative distribution is represented in grey.}
    \label{fig:wl_distrib_runtime_size}
\end{figure*}

\label{subsec:baseline}
The baseline workload is adapted from the most recent available workload
on the Parallel Workload Archive 
KIT-FH2-2016\footnote{\url{https://www.cs.huji.ac.il/labs/parallel/workload/l_kit_fh2/index.html}} log, recorded from the ForHLR II system located at the Karlsruhe Institute of Technology in Germany. 
Notice that this is the most recent workload publicly available and it is still a representative example of a general-purpose computational partition, unaffected by interference from deep learning jobs, better suited for accelerated partitions.
To adapt the log to our testbed, we perform the following steps:

\begin{enumerate}[wide]
    \item The original infrastructure has two queues: 
    the default queue with 1,152 20-core compute nodes, 
    and the visualization queue with 21 nodes comprising CPUs and GPUs.
    We focus on the first queue and exclude the GPU-enabled.

    \item The record is 1.5 years long, and we only have two days of computation available in our testbed. Consequently, we decide to speed up the replay by \textbf{applying a time-scaling factor of 10}, i.e., execution times and inter-arrival times are divided by 10 in the replay\footnote{This factor was chosen to balance the tradeoff between replaying more days from the original workload and suffering from overheads of phenomena that we cannot speed up (like the time taken by scheduling decisions).}.
    In this regard, we can replay $2\times10=20$ days from the original workload.

    \item 
    By studying the distribution of job submissions over time, we notice that the month of July 2017 features a homogeneous volume of submissions over days, well-distributed among users, and the presence of the characteristic day/night and weekday/weekend patterns.
    \review{
    Moreover, the load submitted during this month (611k node-hours) is representative of the other months ($min=199k$, $mean=510k$, $med=576k$, $max=709k$).
    }
    As a result, we select for the experiments \textbf{the 19-day period from July \nth{3} to July \nth{21} (included), 2017}.

    \item Finally, we \textbf{apply the user sampling method} described in Section~\ref{subsec:sampling}, 
    with a target in average node-hour per day of $.84M$, 
    where $M=124\times24$ is the maximum number of node-hour available per day in our testbed.
\end{enumerate}

The resulting workload contains 23 users \review{and 1,895 jobs}.
The median number of node-hours submitted per day is 2,272, i.e., 76\%  of the maximum capacity of the testbed.
The average number of node-hours submitted per day is 2,496, i.e., 84\% of the maximum capacity of the testbed.
Some characteristics of this workload are provided in the form of graphs.
Figure~\ref{fig:wl_distrib_sub} displays the number of daily submissions and node-hours submitted per day. 
Each color represents a particular user, and the stacked graphs show the submission volume per day.
Figure~\ref{fig:wl_distrib_cum} shows the cumulative proportion of jobs as a function of their duration. It shows that a vast majority lasts less than one hour (slightly more than 60\%).
Figure~\ref{fig:wl_distrib_runtime_size} shows the distribution of job execution times and job sizes. 
Jobs are mainly requesting 1, 3, or 4 nodes (see Figure~\ref{fig:wl_distrib_runtime_size_top}). 
But it also shows that large jobs (see 26, 30, and 32 nodes in the x-axis of Figure~\ref{fig:wl_distrib_runtime_size_bottom}), while rare, represent a significant part of the total executed mass.
Since the jobs in the log have a fixed duration, the replayed jobs submitted to the cluster will allocate resources for the same period, either by using \texttt{sleep} commands or implementing active waits.

\textbf{Warm-up Period.}
Looking at Figure~\ref{fig:wl_distrib_sub_bottom}, we note that the first day of the workload, July \nth{3}, features high activity in node-hours submitted.
This day, Monday, comes after a weekend of low activity on the platform. 
However, it is not until 11:00 AM that the platform runs in nominal conditions.
In other words, at that time, resources are virtually fully allocated, and there are pending jobs waiting for resources.
We thus define this period as a ``warm-up'' in the experiments, 
and malleable jobs start being submitted just after it ($t_0$ input of the \textit{generative user} described in Section~\ref{subsec:submitter}).

\subsection{A Generative User: The PhD Student}
\label{subsec:phdstudent}
To thoroughly evaluate our methodology, we count on an additional generative user to submit jobs on top of the baseline workload.
``The student'' is working with a simulation tool, particularly, a positive definite transport equation solver named MPDATA%
\footnote{
The MPDATA algorithm serves as the foundation of the EULAG multiscale fluid solver (Eulerian/semi-Lagrangian)~\cite{rojek_parallelization_2015}, which is responsible for calculating the advection of a non-diffusive quantity in a flow field.
}.
Its algorithm performs iterative time steps to simulate physical phenomena, requiring five input arrays and producing a single output array essential for subsequent time steps. 
The malleable version of MPDATA\footnote{Code available at \url{https://gitlab.bsc.es/siserte/mpdata-dmr}.} redistributes the input arrays between source and target processes, respectively, the number of processes before and after a reconfiguration.
MPDATA malleable demonstrated to increase the utilization of resources and reduce the power consumption in dynamic workloads~\cite{iserte_study_2020}.

For her study, the student must run ten instances ($N$) of MPDATA sequentially.
She will start launching jobs after the warm-up period and wait two scaled hours (720 actual seconds) before submitting the next job. 
This think time ($\Delta t$) corresponds to a hypothetical analysis of the results and the preparation of the subsequent execution.

The student is requested to complete her workload in 14 days; in other words, she has a deadline of two weeks to present her results.
Since she is in a hurry, an alarm is triggered on her phone when a job finishes, and she immediately starts processing data and setting up the next job.

\begin{figure*}[tbp]
    \centering
\includegraphics[clip,width=0.8\linewidth,trim={0.1cm 0.5cm 0.1cm 0.1cm}]{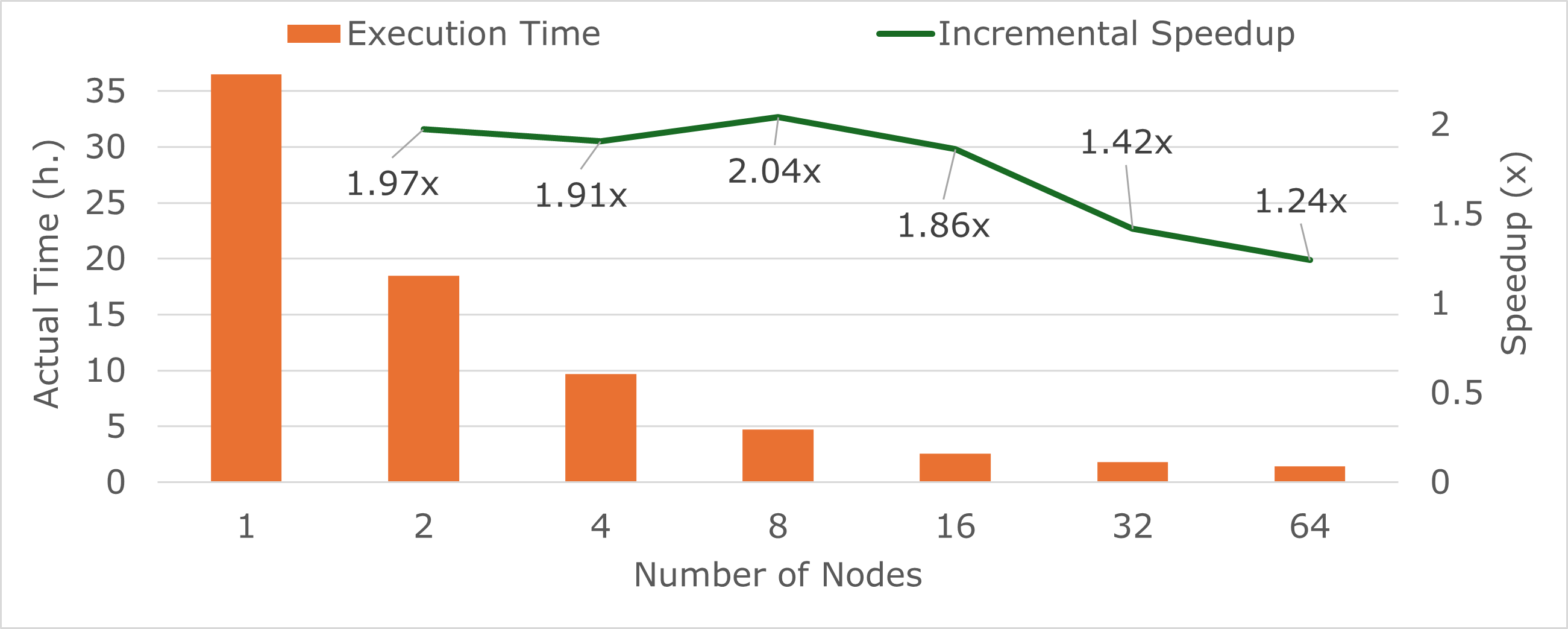}
    \caption{MPDATA scalability in MN5.}
    \label{fig:mpdata-speedup}
\end{figure*}

MPDATA is configured with a computational domain of $8,192\times 1,024\times 128$ cells, iterating during 1,800 steps. 
This computational domain enables the user to run MPDATA from $1$ to $64$ nodes.
Figure~\ref{fig:mpdata-speedup} showcases the scalability of MPDATA with the given configuration in MN5.
The application scales up to $16$ nodes linearly, from where it shows speedups lower than $2x$.

\begin{figure*}[tbp]
    \centering
    \includegraphics[clip,width=\linewidth,trim={0.5cm 0.25cm 0.1cm 0.2cm}]{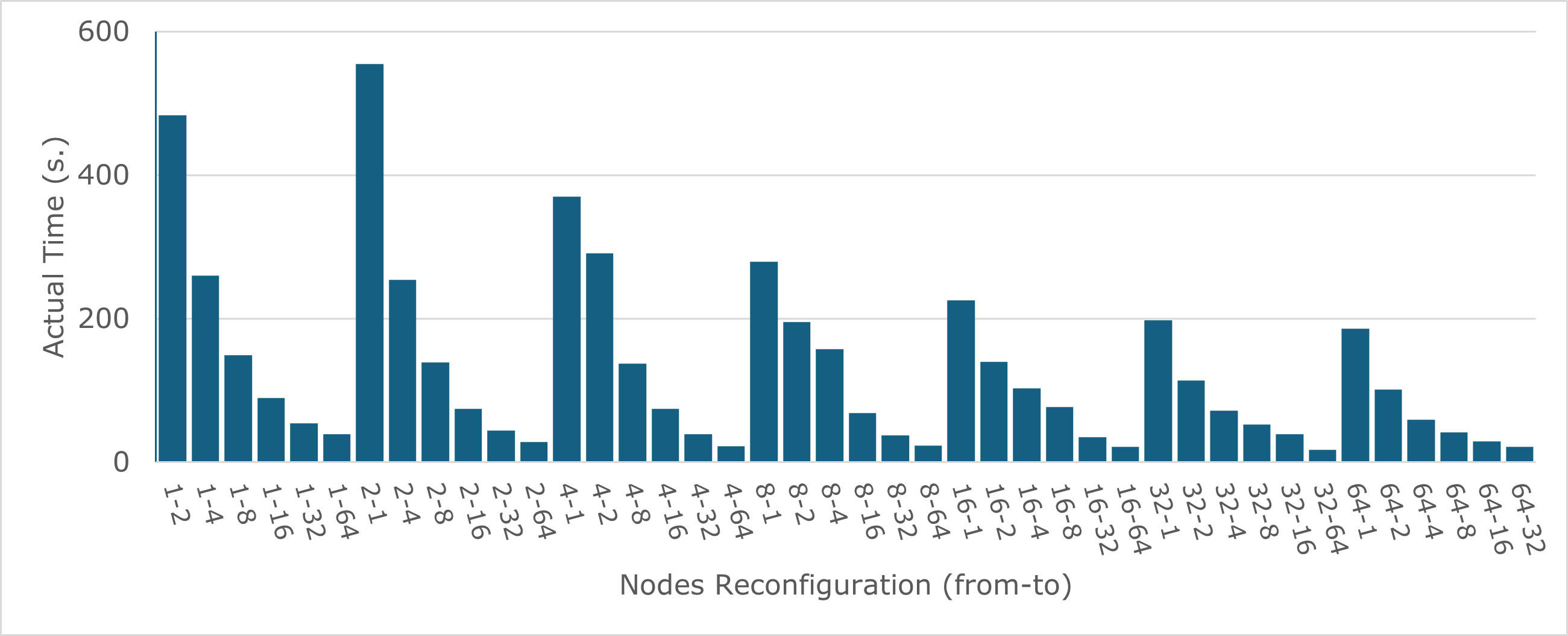}
    \caption{MPDATA reconfiguration times in MN5.}
    \label{fig:mpdata-reconf}
\end{figure*}

\subsection{Experiments Definition}\label{subsubsec:exps}
Five experiments have been designed to study and analyze the effect of job malleability and dynamic management of resources over an existing workload.
Each experiment was launched on an independent 48-hour reservation of the 125-node testbed (see Section~\ref{subsec:testbed}).
\begin{figure*}[tbp]
  \centering
  \begin{minipage}{.6\linewidth}
    \begin{algorithm}[H]
        \caption{Reconfiguration Policy Algorithm in Slurm}
        \label{alg:reconf_policy}
        \begin{algorithmic}[1]
            \STATE $\text{result} \gets \textsc{NONE}$
            \STATE $\eta \gets \textsc{EvaluateMyParallelEfficiency}()$
            \IF{$\textsc{JobCanBeInitiatedWithPartOfMyResources}()$}    
                \IF{$\eta < 0.85$}
                    \STATE $\textsc{SetMaximumPriorityToTargetPendingJob}()$
                     \STATE $\text{result} \gets \textsc{SHRINK}$
                \ENDIF
            \ELSE
                \STATE $\mathcal{R} \gets \textsc{ThereAreAvailableResources}()$
                \IF{$\mathcal{R} \neq \emptyset$}
                    \IF{$\eta > 0.10$}
                        \STATE $\text{result} \gets \textsc{EXPAND}$
                    \ENDIF
                \ENDIF
            \ENDIF
            \STATE \textbf{return} $\text{result}$
        \end{algorithmic}
    \end{algorithm}
  \end{minipage}
\end{figure*}
With these experiments, we will illustrate scenarios that represent different user behaviors:
\begin{enumerate}[wide]
    \item \textbf{Baseline}: this experiment corresponds to executing the baseline workload described in Section~\ref{subsec:baseline}. This experiment sets the original activity of the workload execution in the cluster.
    \item \textbf{StaticN32}: this experiment executes sequentially over the baseline workload, the 10-job workload of the PhD student launched with 32 nodes each.
    This is a realistic case, since with 32 nodes the execution time is $\approx14$ scaled hours (10x the actual time in Figure~\ref{fig:mpdata-speedup}). It means that she could finish her executions in less than six scaled days in an ideal scenario where jobs were not delayed.
    \item \textbf{StaticN16}: this is another non-malleable experiment similar to \textit{Static32} but launching the jobs with 16 nodes, instead.
    In this case, the student works with the shortest configuration time, the one within the 2x speedup in Figure~\ref{fig:mpdata-speedup}, which lasts $\approx18.5$ scaled hours.
    Theoretically, she could complete her executions in less than eight scaled days.
    We have discarded to configure a hypothetical \textit{Static8} experiment, since the theoretical time required to run the 10-job workload with eight nodes under ideal conditions is 47.5 scaled hours, while the maximum wall time granted for the experiment is 48 hours.
    \item \textbf{AlwaysGrow}: in this experiment, the PhD student instantiates malleable jobs over the baseline workload. Jobs are submitted moldable, requesting a 1--64 range of nodes. 
    Since she is under a tight deadline, she decides to disable shrinkages; jobs may only be expanded up to 64 nodes upon the RMS decision. By default, the reconfiguration policy will plan an expansion when sufficient resources are available to upscale.
    Furthermore, the student has defined a reconfiguration inhibitor of \texttt{\#current\_nodes} iterations to avoid abuse of reconfiguration operations and reduce overhead generated by them (see Figure~\ref{fig:mpdata-reconf}).
    \item \textbf{ParEfficiency}: in this case, jobs are submitted within the 2--64 nodes range.
    Malleability limits are defined as one node at the minimum and 64 at the maximum.
    However, the student configures an additional inhibitor for reconfigurations longer than 50 scaled minutes. 
    According to Figure~\ref{fig:mpdata-reconf}, which depicts the measured reconfiguration time for all the possible node combinations of \textit{from--to} reconfigurations, those greater than five minutes are avoided, specifically the reconfigurations \textit{1 to 2}, \textit{2 to 1}, and \textit{4 to 1}.
    This experiment leverages the performance-aware support of the dynamic resource manager (see Section~\ref{subsec:dmr}). 
    Algorithm~\ref{alg:reconf_policy} depicts the reconfiguration policy.
    Every time a malleable job triggers a reconfiguration, Slurm will determine the action that the DMRlib runtime must perform.
    In this regard, Slurm's plugin will check if a pending job may be initiated with some of the resources that would be available if the malleable job relinquishes them after its shrinkage (line~3).
    If a target job may be initiated, the policy checks the current parallel efficiency of the malleable job.
    If the parallel efficiency does not reach a minimum threshold (line~4), the target job priority in the queue will be increased (line~5), and the malleable job will be shrunk (line~6).
    If no job may be initiated and there are available resources in the cluster (line~10), the policy checks the current parallel efficiency. If the value exceeds a determined threshold (line~11), indicating that the execution may still leverage additional resources, the malleable job will be expanded (line~12).

    This policy aims to obtain as many resources as possible if no other job in the queue may use them. However, suppose any job in the queue may be initiated, and the malleable job is not reasonably using the resources (within a parallel efficiency threshold). In that case, the RMS reassigns the resources to a pending job in order to increase global cluster productivity and efficiency.

    \review{
    Notice that preliminary tests varying the PE upper threshold by $\pm(0.05\text{--}0.1)$ and lower threshold by $\pm0.05$ showed no significant deviation in the key outcomes: the student job’s makespan remained consistently reduced, and baseline waiting-time impact was negligible, suggesting that the conclusions are robust within a reasonable range of parameter variation.
    \\\indent
    Furthermore, we completely rely on Slurm’s default scheduling mechanisms, namely the \texttt{sched/backfill} and \texttt{priority/multifactor} policies. When the priority of a job is increased (as in Algorithm~\ref{alg:reconf_policy}, line 5), Slurm recomputes the priorities of all queued jobs, ensuring fairness according to its standard policies. This adjustment may affect the ordering of jobs across the entire queue, not just within feasible backfill windows. However, a detailed design of scheduling and priority policies is beyond the scope of this work, as our focus is specifically on the resource selection policy \texttt{select/linear} that enables malleability.
    }
\end{enumerate}

\begin{table*}[tbp]
    \fontsize{9.5pt}{20pt}\selectfont
    \centering
    \begin{tabular}{lcccccc}
        \toprule
         & & \textbf{Baseline} & \textbf{StaticN32} & \textbf{StaticN16} & \textbf{AlwaysGrow} & \textbf{ParEfficiency} \\
        \midrule
        \multirow{2}{*}{\textbf{Workload Maskespan}} 
            & \textbf{Complete}     & \green{44.84 h.}      & \red{45.15 h.}   & \green{44.84 h.}   & \green{44.84 h.}   & \green{44.84 h.} \\
            & \textbf{Only PhD}     & -             & \red{45.15 h.}   & 38.50 h.   & 36.62 h.   & \green{32.88 h.} \\
        \midrule
        \multirow{2}{*}{\textbf{Global Resource Allocation}} 
            & \textbf{Avg}     & \green{83.70\%}     & \red{93.86\%}  & 91.50\%   & 92.63\%   & 91.86\% \\
            & \textbf{Std}    & 24.07\%     & 11.44\%   & 17.62\%   & 17.24\%   & 17.16\% \\
        \midrule
        \multirow{2}{*}{\textbf{Actual Jobs Waiting Time}}
            & \textbf{Avg}     & \green{1,725.04 s.}    & 3,007.24 s.     & 3,290.38 s.    & \red{4,365.89 s.}     & 4,242.71 s.         \\
            & \textbf{Std}     & 4,450.38 s.    & 6,197.19  s.    & 7,677.89 s.    & 10,013.08 s.    & 11,690.31 s.        \\
        \bottomrule
    \end{tabular}
    \caption{Workloads makespan and average and standard deviation of allocated resources and waiting time for the five experiments, excluding the warm-up period.}
    \label{tab:alloc-wait}
\end{table*}

\begin{figure*}[tbp]
\centering
    \begin{overpic}[clip, width=0.95\linewidth, trim={0.1cm 0.1cm 0.1cm 0.2cm}]{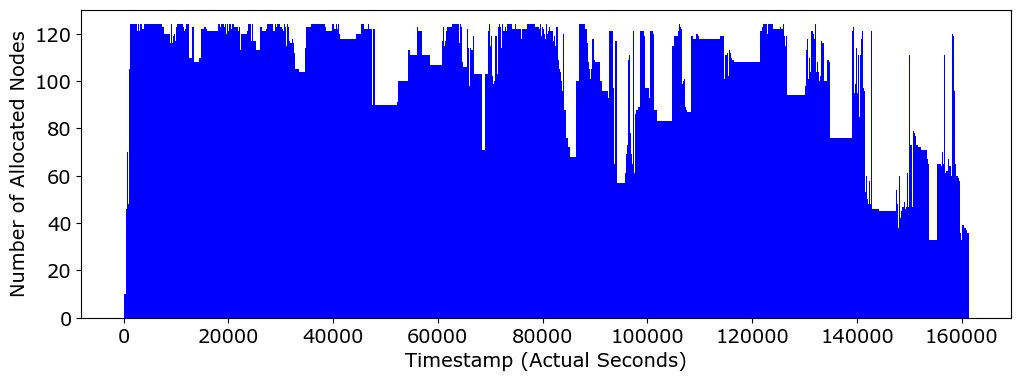}
    \end{overpic}
    \caption{Resource allocation (Y-axis) in each second of the execution (X-axis) for \textit{Baseline} experiment.}
    \label{fig:baseline}
\end{figure*}

\section{Results}
\label{sec:results}
In this section, the results obtained after running the experiments are analyzed.
We use the following metrics for the validation of MPI malleability: makespan (Section~\ref{subsec:makespan}), resource allocation rate (Section~\ref{subsec:utilization}), and waiting time (Section~\ref{subsec:wt})%
\footnote{
Since the execution time is fixed and determined by the log, and the completion time is the sum of waiting plus execution times, the completion time is entirely dependent on the waiting time.
}.

Note that the metrics have different interpretations depending on the type of workload replay, as well-explained by Feitelson~\cite{feitelson_resampling_2021}.
In the case of replay with feedback (like here with the PhD student workload), the primary performance metric is the makespan, i.e., the total time to execute the workload.
On the contrary, the makespan in timestamp-based replay (like here with the baseline workload) is dictated by the original submission timestamps.
In this case, the primary performance metric is the waiting times for each job.

\begin{figure*}[tbp]
    \centering

    \begin{subfigure}[t]{0.95\textwidth}
        \centering
        \begin{overpic}[clip, width=\linewidth, trim={0.1cm 1.4cm 0.1cm 0.2cm}]{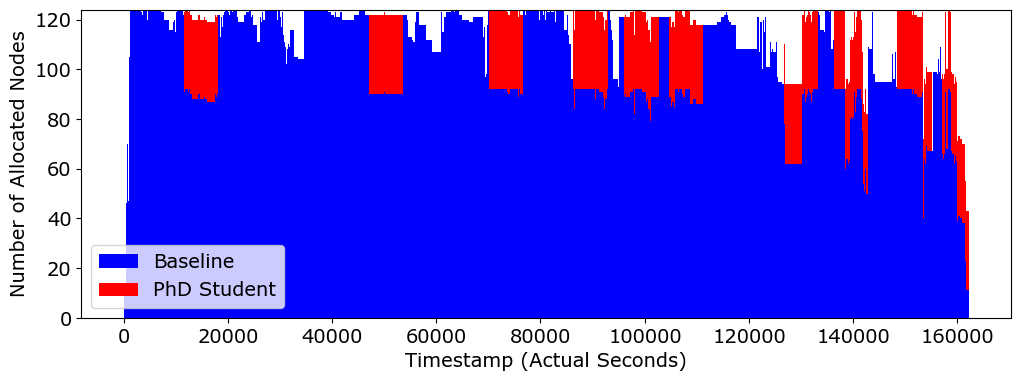}
            \put(73.5,0){\color{gray}\line(0,1){32}}
            \put(74.55,10){\color{gray}\small Deadline}
        \end{overpic}
        \caption{\textit{StaticN32}}
        \label{multifig:static32}
        \vspace{3mm}
    \end{subfigure}

    \begin{subfigure}[t]{0.95\textwidth}
        \centering
        \begin{overpic}[clip, width=\linewidth, trim={0.1cm 0.1cm 0.1cm 0cm}]{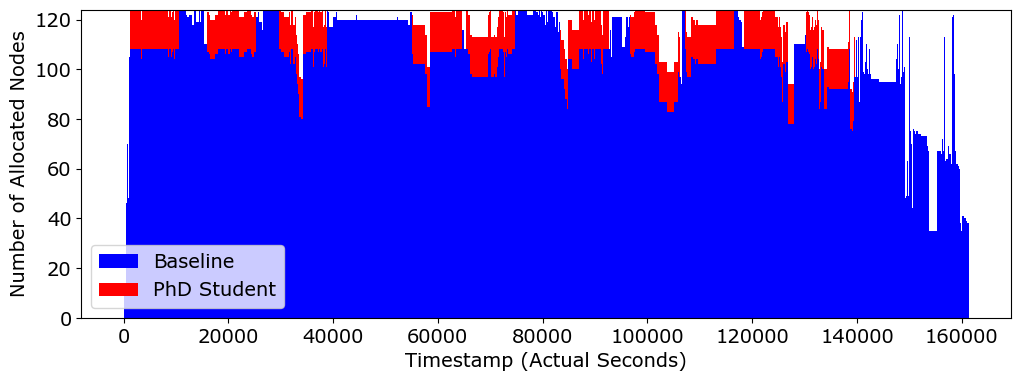}
            \put(73.95,4){\color{gray}\line(0,1){33}}
            \put(75,10){\color{gray}\small Deadline}
        \end{overpic}

        \caption{\textit{StaticN16}}
        \label{multifig:static16}
    \end{subfigure}

    \caption{
    Resource allocation (Y-axis) over execution time (X-axis) for the static experiments of the student.} 
    \label{fig:static-experiments}
\end{figure*}

\begin{figure*}[tbp]
    \centering

    \begin{subfigure}[t]{0.95\textwidth}
        \centering
        \begin{overpic}[clip, width=\linewidth, trim={0.1cm 1.4cm 0.1cm 0.2cm}]{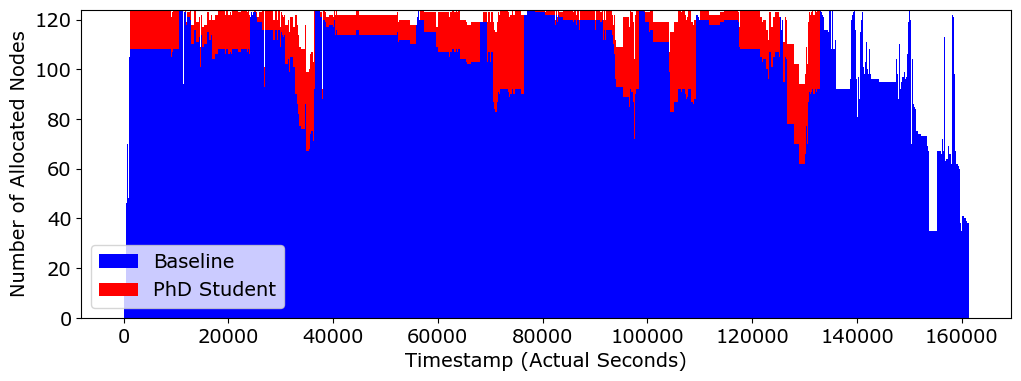}
            \put(73.95,0){\color{gray}\line(0,1){32}}
            \put(75,10){\color{gray}\small Deadline}
        \end{overpic}

        \caption{\textit{AlwaysGrow}}
        \label{multifig:dyngrow}
        \vspace{3mm}
    \end{subfigure}

    \begin{subfigure}[t]{0.95\textwidth}
        \centering
        \begin{overpic}[clip, width=\linewidth, trim={0.1cm 0.1cm 0.1cm 0cm}]{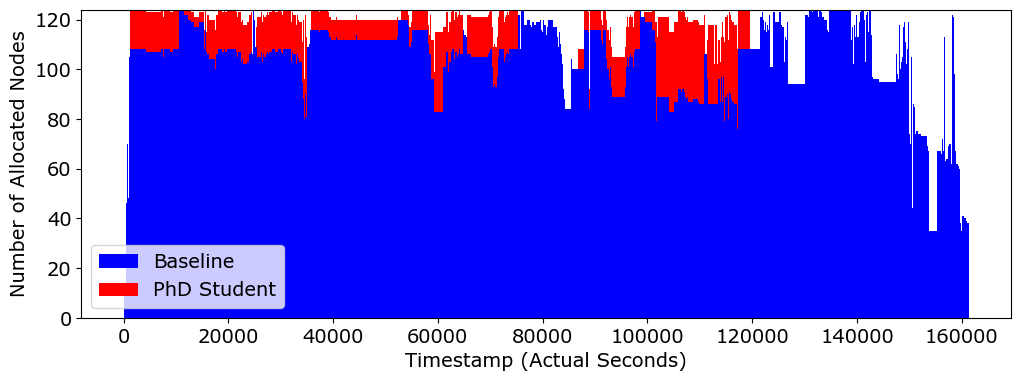}
            \put(73.95,4){\color{gray}\line(0,1){33}}
            \put(75,10){\color{gray}\small Deadline}
        \end{overpic}
        
        \caption{\textit{ParEfficiency}}
        \label{multifig:dynpe}
    \end{subfigure}

    \caption{
    Resource allocation (Y-axis) over execution time (X-axis) for the dynamic experiments of the student.} 
    \label{fig:dynamic-experiments}
\end{figure*}

\subsection{Makespan}\label{subsec:makespan}
\review{
The completion time of the workload, or workload makespan, is a critical metric for the PhD student, who has a deadline to meet.
Particularly, she needs to focus on the time it takes to complete her 10-job workload.
Table~\ref{tab:alloc-wait} compiles execution metrics of the different experiments.
The first two rows of Table~\ref{tab:alloc-wait} (after the header) report the execution times for the two workloads. The first row corresponds to the complete workload (Baseline + PhD student), which also defines the overall experiment duration, while the second row refers exclusively to the PhD student workload.
\\\indent
As discussed in the introduction of this section, the first row is not meaningful, since makespan is dictated by the rigid submission timestamps replayed from the original log.
This is why all the values in this row are equal to the \textit{Baseline} makespan.
The only exception is \textit{StaticN32}, whose makespan is ruled by the PhD job, as shown in Figure~\ref{multifig:static32}.
The difficulty of scheduling large jobs in a congested system is manisfested in this experiment that reports a completion time increased compared to \textit{Baseline}.
\\\indent
In the remaining experiments (\textit{StaticN16}, \textit{AlwaysGrow}, and \textit{ParEfficiency}), the PhD student jobs are completed before the end of the baseline workflow.
Of course, reducing the PhD student jobs size from 32 to 16 nodes helps the backfilling scheduler to find gaps to fit the jobs, thus reducing the PhD student workload completion time from 45.15 to 38.50 actual hours.
However, it is not enough to meet the student's deadline of 14 scaled days.
Figure~\ref{multifig:static16} shows that the last two jobs of the student are still running after the deadline.
\textit{AlwaysGrow}, leverages malleability to reduce the student makespan, but it still needs 36.62 actual hours, which is still beyond the deadline (see Figure~\ref{multifig:dyngrow}).
Notably, the ``smartest'' dynamic configuration, \textit{ParEfficiency}, needs $\approx73\%$ the time to complete the student jobs compared to \textit{StaticN32} (32.88 h. vs 45.15 h., respectively) without delaying the baseline workload and meeting the deadline, since both full workloads are completed in 44.84 h.
To sum up, Figure~\ref{multifig:dynpe} shows that the last job of the student finishes before the deadline in \textit{ParEfficiency} scenario.
\\\indent
In the following subsections, we will see how malleability affects the resource allocation rate and the waiting time, which are the two factors that explain the makespan reduction of the PhD student workload.
}

\subsection{Resource Allocation Rate}\label{subsec:utilization}
Figure~\ref{fig:baseline} showcases the total number of allocated nodes for every actual second of the \textit{Baseline} experiment.
Correspondingly, figures~\ref{multifig:static32},~\ref{multifig:static16},~\ref{multifig:dyngrow}, and~\ref{multifig:dynpe} showcase stacked bar plots where each timestamp depicts the sum of the allocated resources by the baseline plus the PhD student workloads.
The figures present different patterns: while the static experiments in figures~\ref{multifig:static32} and~\ref{multifig:static16} show how the student jobs have to wait for slots that satisfy the static requests, the malleable experiments in figures~\ref{multifig:dyngrow} and~\ref{multifig:dynpe} leverage fragmentation where student jobs run.

\review{
Table~\ref{tab:alloc-wait} contains in the \nth{3} row the allocation rate for each experiment, which is the average node utilization over the makespan of the experiment, excluding the warm-up period.
In turn, the \nth{4} row presents the standard deviation of the allocation rate.
\\\indent
\textit{Baseline} provides the lower average allocation rate with the highest standard deviation, which is expected since the platform presents many idle regions of resources during the execution (see Figure~\ref{fig:baseline}).
Evidently, the remaining experiments present higher allocation rates because of the 10 extra jobs of the PhD student.
While \textit{StaticN32} presents the highest average allocation rate, it also shows the lowest standard deviation, which means that the platform is uniformly more saturated during the experiment (see Figure~\ref{multifig:static32}).
The remaining experiments present similar average allocation rates, with higher standard deviations than \textit{StaticN32}, which means that the platform presents more idle regions during the execution (see figures~\ref{multifig:static16},~\ref{multifig:dyngrow}, and~\ref{multifig:dynpe}), particularly at the end of the workload execution, when the jobs of the baseline workload are still running, but the PhD student jobs have already finished.
In this regard, the rigid jobs of \textit{StaticN16} and the malleable jobs of \textit{AlwaysGrow} and \textit{ParEfficiency} make no difference in the allocation rate.
That is why we need to understand better the makespan reduction of the PhD student workload with studying the waiting time in the next section, where we see how MPI malleability can maintain resource utilization while reducing the completion times.
}

\subsection{Waiting time}\label{subsec:wt}
\review{
To analyze how baseline jobs are affected by the submission of jobs from the PhD student, we look at their average waiting times and standard deviation in the different scenarios (see Table~\ref{tab:alloc-wait}, \nth{5} and \nth{6} rows).
\\\indent
It is patent that the original workload suffers longer delays on average when the PhD student jobs are submitted: the average waiting time increases from 1,725.04 s. to 3,007.24 s. or 4,365.89 s. in the best and worst cases, respectively.
This increment of up to $\approx2.53x$ is expected since the system is more saturated with the extra jobs.
\\\indent
Besides, the two dynamic experiments (\textit{AlwaysGrow} and \textit{ParEfficiency}) pose a larger impact on the baseline waiting time, in terms of average and standard deviation.
This can be explained by the greedy nature of the malleable configurations, starting with fewer resources and expanding as resources are released in the platform, making the baseline jobs wait longer.
}

\begin{figure*}[tbp]
    \centering
    \begin{overpic}[clip,width=0.95\linewidth,trim={0.1cm 0.1cm 0.1cm 0.1cm}]{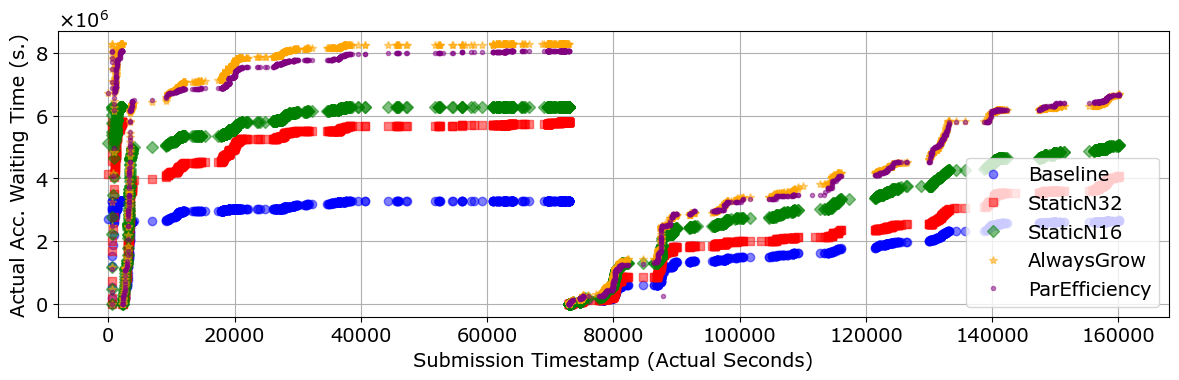}
        \put(48.4,4){\color{gray}\line(0,1){26}} 
        \put(49.5,18){\color{gray}\small July \nth{10}} 
    \end{overpic}
    \caption{Accumulated waiting time throughout the workloads executions.}
    \label{fig:waitAcc}
\end{figure*}

\begin{figure*}[tbp]
    \centering
    \begin{overpic}[clip,width=0.95\linewidth,trim={0.1cm 0.1cm 0.1cm 0.1cm}]{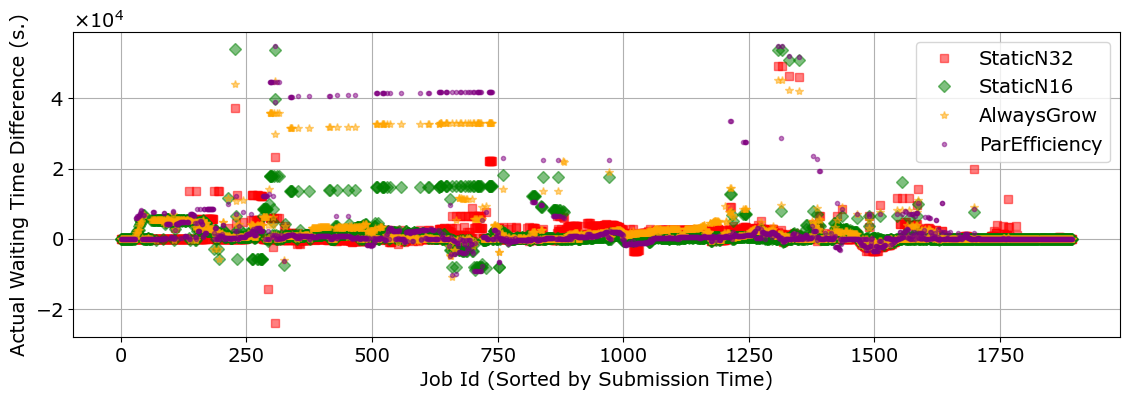}
        \put(44.8,4){\color{gray}\line(0,1){29}} 
        \put(45.8,24){\color{gray}\small July \nth{10}} 
    \end{overpic}
    \caption{Difference in job waiting times compared to the baseline experiment.} 
    \label{fig:waitDiff}
\end{figure*}

In a finer detailed analysis, we studied the accumulated waiting time throughout the execution (Figure~\ref{fig:waitAcc}).
It shows how the waiting time increases as new jobs arrive in the queue, and since there are not enough available resources until around timestamp 77,000 (July \nth{10}), they cannot be quickly started.
After this milestone, the waiting time progressively starts growing again. 
More specifically, Figure~\ref{fig:waitDiff} showcases the waiting time difference of the various experiments compared to the Baseline for each job.
One of the most valuable insights we extract is the behavior detected around job $750$ (July \nth{10}), where waiting times drop.
This is because there is a period around July \nth{8} and \nth{9}, as shown in Figure~\ref{fig:wl_distrib_sub_top}, where the submissions drastically decrease, and it takes over a day to drain the queue.
That is why jobs submitted after that event present a virtually null waiting time.

\review{
The gap in think time observed around timestamp 80,000 in Figure~\ref{multifig:dynpe} corresponds to Job~7 (see Table~\ref{tab:reconfig_pairs}), which experienced a delay of nearly four hours before initiation. This behavior is not an artifact of our methodology but rather the result of the scheduling dynamics in Slurm. Specifically, the delay was caused by the higher priority of other pending jobs in the queue. While backfilling can allow jobs to advance when resources are available, Slurm’s fairness policy also incorporates accumulated waiting time into its priority calculation. Consequently, student's Job~7 remained pending until its priority increased sufficiently to be scheduled.}

\subsection{Focus on the PhD Student}
\begin{figure*}[tbp]
    \centering
        \includegraphics[clip,width=0.95\linewidth,trim={0.1cm 0.1cm 0.1cm 0.1cm}]{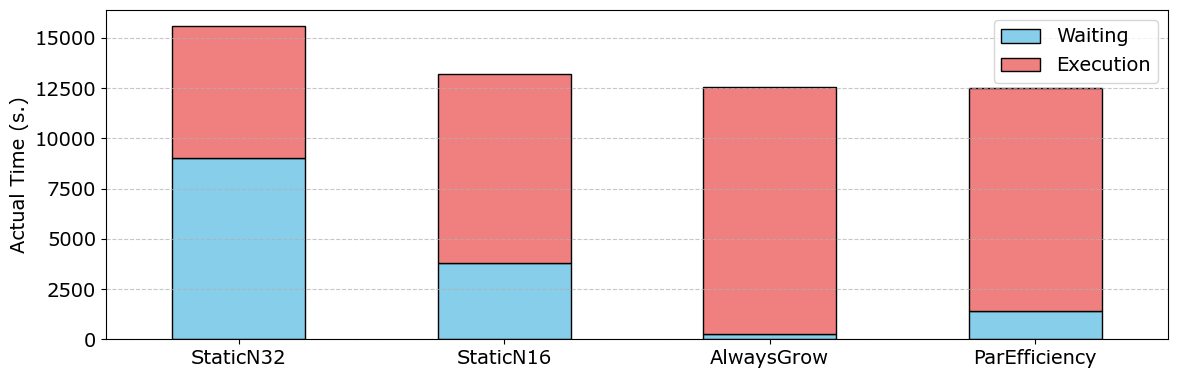}
        \caption{Average student's job completion (waiting + execution) time (Y-axis) for the four experiments (X-axis).}
        \label{fig:phd_completion}
\end{figure*}

\review{
Figure~\ref{fig:phd_completion} illustrates the average completion time of the student's jobs, defined as the sum of waiting and execution times. 
Dynamic experiments exhibit longer execution times than the static ones. 
This is because malleable jobs are submitted with a range of possible node counts, and the RMS can more easily allocate resources within that range at submission time, typically closer to the lower bound. 
By leveraging backfilling, these smaller jobs start earlier, so the \textit{Student}'s jobs experience almost no waiting time when malleability is enabled.
\\\indent
Once started, malleable jobs are progressively expanded according to their respective reconfiguration policies. 
Assigning more resources reduces execution time (see Figure~\ref{fig:mpdata-speedup}), but each reconfiguration incurs an overhead (see Figure~\ref{fig:mpdata-reconf}), which is accounted for in the execution time.
}

\begin{figure*}[tbp]
    \centering
        \includegraphics[clip,width=0.95\linewidth,trim={0.2cm 0.1cm 0.2cm 0.1cm}]{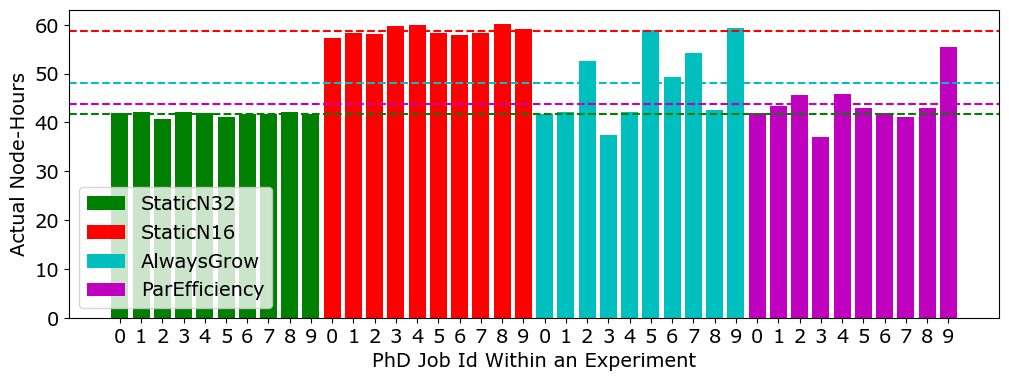}
        \caption{Individual student job's Node-hours (Y-axis) grouped by colors for the four experiments (X-axis). Dashed lines represent the average per experiment (Y-axis).}
        \label{fig:phd_nodehours}
\end{figure*}

\begin{table*}[tbp]
\centering
\fontsize{9.5pt}{14pt}\selectfont
\begin{tabular}{l r r r  r r r  r r r}
    \toprule
    & \multicolumn{3}{c}{Expansions} & \multicolumn{3}{c}{Shrinkages} & \multicolumn{3}{c}{Total} \\
    \cmidrule(lr){2-4} \cmidrule(lr){5-7} \cmidrule(lr){8-10}
    & \# & Avg. & Overhead & \# & Avg. & Overhead & \# & Avg. & Overhead \\
    \midrule
    AlwaysGrow   & 27 & 160 s. & 4,332 s. & 0  & --    & 0 s.    & 27 & 160 s. & 4,332 s. \\
    ParEfficiency & 18 &  87.8 s.  & 1,580 s. & 7  & 48.9 s. &  349 s.  & 25 & 76.9 s.  & 1,929 s. \\
    \bottomrule
\end{tabular}
\caption{Per experiment: number of reconfigurations (expansions / shrinkages), average reconfiguration time and accumulated overhead (actual seconds).}
\label{tab:reconfig}
\end{table*}

\review{
The total cost of reconfigurations is summarized in Table~\ref{tab:reconfig}. 
\\\indent
The \textit{AlwaysGrow} policy was designed to aggressively exploit every opportunity for expansion, regardless of the immediate overhead, with the goal of reaching the job’s maximum resource allocation as quickly as possible. Although some of these reconfigurations are slow and yield limited benefits, they highlight the trade-offs of prioritizing rapid growth over efficiency. 
\\\indent
In contrast, the \textit{ParEfficiency} policy applies a more selective approach, avoiding the most expensive reconfigurations and thereby achieving better overall efficiency. The juxtaposition of these two strategies is deliberate: \textit{AlwaysGrow} illustrates the risks of unfiltered expansion, whereas \textit{ParEfficiency} demonstrates how threshold-based inhibition can achieve more balanced outcomes. Together, they underscore the need for adaptive policies and justify the design of tunable malleability mechanisms in HPC environments. For completeness, the detailed reconfiguration events of the \textit{ParEfficiency} experiment are listed in Table~\ref{tab:reconfig_pairs}.
\\\indent
A key advantage of \textit{ParEfficiency} is its ability to reduce overhead by preventing the slowest reconfigurations. Moreover, this policy adapts malleable jobs to workload and performance dynamics, triggering seven \textit{32-to-16} shrinkages. These shrinkages improve cluster productivity by reducing the PhD student workload time by 10\% compared to \textit{AlwaysGrow}.
\\\indent
The impact of these policies is also reflected in node-hour consumption. Figure~\ref{fig:phd_nodehours} reports the node-hours consumed by PhD student jobs across all experiments, with the dashed line showing the mean per configuration. Static jobs display a virtually constant node-hour consumption (standard deviation $<$ $1$). In an ideal linear-scaling scenario (i.e., a $2\times$ speedup when doubling the number of nodes), the node-hour consumption would be identical in \textit{StaticN32} and \textit{StaticN16}. However, the application exhibits sublinear scalability (see Figure~\ref{fig:mpdata-speedup}), achieving only a $1.42\times$ speedup when increasing from 16 to 32 nodes. As a result, the \textit{StaticN32} configuration consumes 17 additional node-hours compared to \textit{StaticN16}.
\\\indent
Malleable jobs, by design, show variable node-hour consumption within the experiments (standard deviation $>$ $4$). Interestingly, both \textit{AlwaysGrow} and \textit{ParEfficiency} contain the two ``cheapest'' PhD student jobs, which spend most—or even all—of their execution on 8 nodes. This is significant, as eight nodes provide the best efficiency point for this application, leading to the lowest node-hour cost.
}

\begin{table}
\caption{Reconfigurations in the \textit{ParEfficiency} experiment (time in actual seconds). Gray rows represent shrinkages.}
\label{tab:reconfig_pairs}
\begin{tabular}{clll}
\toprule
\textbf{PhD Job Id}& \textbf{Initial Nodes} & \textbf{Final Nodes} & \textbf{Rec. Time} \\
\midrule
 \multirow{1}{*}{\textbf{0}} 
            & 16 & -- & -- \\
\midrule
 \multirow{3}{*}{\textbf{1}} 
            & 2 & 4 & 254 s. \\
            & 4 & 8 & 136 s.\\
            & 8 & 16 & 72 s.\\
\midrule
 \multirow{4}{*}{\textbf{2}} 
            & 8 & 16 & 69 s.\\
            & 16 & 32 & 43 s.\\
            & \cellcolor{lightgray}32 & \cellcolor{lightgray}16 & \cellcolor{lightgray}47 s.\\
            & 16 & 32 & 34 s.\\
\midrule
 \multirow{1}{*}{\textbf{3}} 
            & 8 & -- & -- \\
\midrule
 \multirow{4}{*}{\textbf{4}} 
            & 4 & 8 & 133 s.\\
            & 8 & 16 & 69 s.\\
            & 16 & 32 & 35 s.\\
            & \cellcolor{lightgray}32 & \cellcolor{lightgray}16 & \cellcolor{lightgray}49 s.\\
\midrule
 \multirow{1}{*}{\textbf{5}} 
            & 8 & 16 & 68 s.\\
\midrule
\multirow{1}{*}{\textbf{6}} 
            & 8 & 16 & 66 s.\\
\midrule
\multirow{1}{*}{\textbf{7}} 
            & \cellcolor{lightgray}32 & \cellcolor{lightgray}16 & \cellcolor{lightgray}47 s.\\
\midrule
\multirow{3}{*}{\textbf{8}} 
                & 2 & 4 & 253 s.\\
                & 4 & 8 & 135 s.\\
                & 8 & 16 & 67 s.\\
\midrule
 \multirow{8}{*}{\textbf{9}} 
            & 16 & 32 & 38 s.\\
            & \cellcolor{lightgray}32 & \cellcolor{lightgray}16 & \cellcolor{lightgray}48 s.\\
            & 16 & 32 & 36 s.\\
            & \cellcolor{lightgray}32 & \cellcolor{lightgray}16 & \cellcolor{lightgray}51 s.\\
            & 16 & 32 & 38 s.\\
            & \cellcolor{lightgray}32 & \cellcolor{lightgray}16 & \cellcolor{lightgray}51 s.\\
            & 16 & 32 & 34 s.\\
            & \cellcolor{lightgray}32 & \cellcolor{lightgray}16 & \cellcolor{lightgray}50 s.\\        
\bottomrule
\end{tabular}
\end{table}

\section{Discussion}\label{sec:discussion}
Our methodology adapts historical workload traces to a target infrastructure, accounting for the fact that reproducing logs on the same system is often not feasible. The goal is to provide a realistic path for gradual DRM adoption in HPC centers.

We do not expect full adoption of DRM techniques to happen immediately. Instead, early adopters (technically skilled users motivated to exploit malleability) would be the first to submit DRM-enabled jobs to production environments still dominated by fixed-size workloads. This hybrid workload composition enables us to assess the incremental benefits of malleability without needing full-scale user migration.

This setup also offers valuable insight into how malleable jobs can coexist with standard jobs and make use of fragmented resources that would otherwise go idle. Running our experiments on a real Slurm deployment allowed us to combine malleability with Slurm’s native backfilling policy, improving resource utilization.

While the number of test scenarios is limited by the need for live system execution, our methodology produces high-fidelity results. Moreover, it can be reused by organizations with their own workload traces, enabling site-specific evaluation of malleability’s practical benefits and deployment challenges.

We argue that our method is more robust than many alternatives commonly used in the literature (see Section~\ref{sec:related}). We (i) use a real workload trace rather than synthetic jobs, and (ii) right-size the workload by randomly sampling users, which preserves authentic submission and temporal patterns.


The interpretation of our results highlights several important trade-offs.
At the end of the day, the PhD student successfully meets the deadline in the \textit{ParEfficiency} scenario. Counterintuitively, large fixed resource requests, which users might assume to be beneficial, actually work against minimizing completion time. Additionally, overly aggressive malleable strategies can introduce excessive reconfiguration overhead, limiting their ability to efficiently leverage resource fragmentation.  

\begin{table*}[tbp]
    \centering
    \fontsize{9.5pt}{20pt}\selectfont
    \begin{tabular}{lcccc}
        \toprule
        & \textbf{StaticN32} & \textbf{StaticN16} & \textbf{AlwaysGrow} & \textbf{ParEfficiency} \\
        \midrule
        \textbf{Total}    & 232.83 n/day (100\%)  & 200.54 n/day (100\%)  & 189.09 n/day (100\%)  & 171.52 n/day (100\%)  \\
        \textbf{Baseline} & 193.5 n/day (83.11\%) & 173.7 n/day (86.62\%) & 164.2 n/day (86.84\%) & 145.68 n/day (84.93\%) \\
        \textbf{Student}  & 23.95 n/day (10.29\%) & 17.4 n/day (8.68\%)   & 20.01 n/day (10.58\%) & 18.25 n/day (10.64\%)  \\
        \textbf{Accumulated} & 217.45 n/day (93.39\%)& 191.1 n/day (95.29\%) & 184.21 n/day (97.42\%)& 163.93 n/day (95.57\%) \\
        \bottomrule
    \end{tabular}
    \caption{Resource consumption (nodes/day) and its total percentage during the student's activity for the different experiments.}
    \label{tab:nodes-day}
\end{table*}

Fragmented resource availability, a common characteristic of HPC infrastructures, represents the ideal use case where malleability becomes crucial. 
This fragmentation is what our student leverages.
Table~\ref{tab:nodes-day} focuses on the time period during which the student is actively running jobs (corresponding to the red areas in figures~\ref{multifig:static32},~\ref{multifig:static16},~\ref{multifig:dyngrow}, and~\ref{multifig:dynpe}). The table presents the resource consumption in \textit{nodes/day} for the workloads, along with their proportion of the total available area ($124$ nodes $\times$ time) shown in the first row.  

The data highlights an interesting trade-off. Although the most cost-effective strategy for the student is \textit{StaticN16} (third row), it does not only allow meeting the deadline but also results in increased costs for the baseline workload (second row) and higher overall cluster resource consumption during that period (fourth row), compared to the optimal \textit{ParEfficiency}.

Given these findings, users who contribute to higher infrastructure efficiency by running \textit{parallel-efficiency-aware malleable jobs}—which increase resource utilization without extending job execution times—could be incentivized. One possible approach would be to award a reduction in their usage quota proportional to their efficiency gains, encouraging broader adoption of malleability in HPC environments.

\review{
Despite our best efforts, we recognize several potential threats to the validity and generalizability of our results. First, baseline jobs were replayed using active waits rather than real computations, which could in principle overlook issues of resource interference. This would be problematic in environments where resources are shared among concurrent jobs. However, in HPC systems such as MareNostrum 5---the target of our work---nodes within large partitions are allocated exclusively to individual jobs, ensuring isolation and preventing interference, thereby safeguarding the integrity of our setup.
\\\indent
Second, the original trace was compressed by a factor of 10 to replay more days within our 48-hour reservations. While this reduces realism by amplifying phenomena that cannot be sped up, such as scheduler decision times, both execution and inter-arrival times were consistently scaled. As a result, queue dynamics and workload arrival patterns remain preserved, and all experiments remain directly comparable.
\\\indent
Third, our evaluation considered only one dynamic job at a time, thus not capturing competition between multiple dynamic jobs. This was an intentional choice to isolate and analyze malleability in controlled conditions. Still, our preliminary work~\cite{iserte_dmrlib_2020} has shown that scenarios with multiple dynamic jobs are feasible and have been explored in earlier studies.
\\\indent
Fourth, we evaluated only a single application, MPDATA. This code was selected because of its step-based structure and near-linear scalability, traits that are representative of many HPC workloads. While our empirical results are application-specific, MPDATA serves as a meaningful abstraction for a broad class of malleable applications. Furthermore, previous work~\cite{iserte_dmrlib_2020} studied additional applications with diverse communication/computation balances, supporting the generality of our methodology.
\\\indent
Finally, experiments were performed using a single input dataset---the July 2017 segment of the KIT-FH2-2016 workload---and a fixed random seed for user sampling, without replication. 
In our experience, and consistent with practices commonly found in the state of the art, execution times at this scale tend to be stable enough that single runs provide representative results, unlike microbenchmarks (e.g., bandwidth or latency) where variability can significantly affect outcomes.
Nevertheless, replicating the study with alternative inputs or complementing it with simulations would strengthen confidence in the conclusions while avoiding the prohibitive costs of large-scale experimental campaigns.
}
\rebuttal{
Nevertheless, we agree that including confidence intervals and statistical analyses would strengthen future work, particularly for smaller-scale or reduced-size workloads where multiple repetitions are feasible.}

\section{Conclusion}\label{sec:conclusion}
In this work, we have introduced a novel methodology to validate dynamic resource management techniques by adapting supercomputer logs to DRM-enabled computing clusters.
Unlike previous studies that relied on benchmarks or simulations with synthetic workloads, we employ a novel methodology called \textit{User-Based Submitter} that combines user sampling and workload replay with feedback.
By leveraging real-world data and testing on an MPI malleability-enabled supercomputer's partition, we provide a realistic and actionable framework for evaluating new resource management and job scheduling policies.

\review{
We demonstrated the effectiveness of our approach through a novel use case, where an HPC user successfully leveraged MPI malleability to complete her workload on time. Validation in a supercomputing environment showed a 27\% reduction in malleable workload time maintaining the resource utilization rate. While individual jobs may experience delays, the baseline workload remains unaffected in terms of execution time. We believe this methodology is broadly applicable and represents a promising step toward integrating dynamic resource management into next-generation HPC infrastructures.
}

\section*{CRediT authorship contribution statement}
\begin{itemize}
    \item \textbf{Sergio Iserte:} Conceptualization, Methodology, Software, Investigation, Writing - Original Draft, Visualization, Project Administration.
    \item \textbf{Maël Madon:}  Methodology, Investigation, Software, Writing - Original Draft, Visualization.
    \item \textbf{Georges Da Costa:} Investigation, Writing - Review \& Editing, Supervision.
    \item \textbf{Jean-Marc Pierson:} Investigation, Writing - Review \& Editing, Supervision.
    \item \textbf{Antonio J. Peña:} Resources, Writing - Review \& Editing, Supervision, Funding Acquisition.
\end{itemize}

\section*{Declaration of competing interest}
The authors have no conflicts of interest to declare that are relevant to the content of this article.

\section*{Funding sources}
The researchers from BSC are involved in the project The European PILOT, which has received funding from the European High-Performance Computing Joint Undertaking (JU) under grant agreements No. 101034126 and No. PCI2021-122090-2A under the MCIN/AEI and the EU NextGenerationEU/PRTR.
Antonio J. Peña was partially supported by the Ramón y Cajal fellowship RYC2020-030054-I funded by MCIN/AEI/ 10.13039/501100011033 and by ``ESF Investing in your future''.

\section*{Data availability}
The artifact source-code can be cloned from GitHub: \url{https://github.com/siserte/dmr-poc}.

\section*{Declaration of generative AI and AI-assisted technologies in the writing process}
During the preparation of this work the authors used ChatGPT in order to improve language and readability. After using this tool/service, the authors reviewed and edited the content as needed and take full responsibility for the content of the publication.

\bibliographystyle{elsarticle-num}
\bibliography{bib}



\section*{Supplementary material}

# Appendix: Artifact Description

**Artifact Description (AD)**

## 1 Overview of Contributions and Artifacts

### 1.1 Paper's Main Contributions

- $C_1$ A methodology to replay HPC logs in any malleability-enabled computational cluster.
- $C_2$ The design and study of a novel case study of a production supercomputer that initiates the adoption of malleable jobs into its traditional rigid workload.
- $C_3$ The validation of MPI malleability using a real workload in a malleability-enabled supercomputer.

### 1.2 Computational Artifacts

The artifact source-code can be cloned from GitHub: https://github.com/siserte/dmr-poc.

- $A_1$ `./dmr.h` and `./dmr.c` files
- $A_2$ `./slurm-spawn/` directory
- $A_3$ `./poc/` directory
- $A_4$ `./mpdata-dmr/` directory
- $A_5$ `./workload-generation/` and `./paper-experiments` directories

| Artifact ID | Contributions Supported | Related Paper Elements |
|---|---|---|
| $A_1$ | $C_2, C_3$ | Figures 5, 7-15<br>Tables 1-2 |
| $A_2$ | $C_2, C_3$ | Algorithm 2 |
| $A_3$ | $C_1, C_3$ | Algorithm 1<br>Figures 1-3 |
| $A_4$ | $C_2$ | Figures 4-5, 7-15 |
| $A_5$ | $C_3$ | Figures 6-15<br>Tables 1-2 |

## 2 Artifact Identification

### 2.1 Computational Artifact $A_1$

**Relation To Contributions**

The artifact $A_1$ contains the Dynamic Management of Resources Library (DMRlib) which is responsible for linking malleable applications with the resource manager and orchestrates resources and MPI processes. In this regard, we leveraged DMRlib for designing the study case ($C_2$) and to validate MPI malleability in a supercomputer ($C_3$).

**Expected Results**

DMRlib automatically reallocates resources, redistributes data, and reshape the MPI process layout of the malleable applications ($A_4$).

**Expected Reproduction Time (in Minutes)**

DMRlib is the framework so the time required to reproduce it corresponds to the setup which takes less than one minute.

**Artifact Setup (incl. Inputs)**

*Hardware.* Any HPC cluster. We need at least 3 nodes to evaluate DMRlib: one management node and two compute nodes to enable expansions to more than one node and shrinkages up to one node.

*Software.*

(1) GCC v10.4+
(2) Although in the paper we leveraged MPICH v3.2 (https://www.mpich.org/static/downloads/3.2/mpich-3.2.tar.gz), OpenMPI v5+ is also supported.
(3) DLB v3.5.0 (https://pm.bsc.es/ftp/dlb/releases/dlb-3.5.0.tar.gz).
(4) DMRlib requires $A_2$.

*Datasets / Inputs.* Users need to define *communication efficiency* targets and *reconfiguration inhibition* periods. DMRlib is invoked by the jobs and the input is described in the experiments of $A_5$.

*Installation and Deployment.* DMRlib can be installed running make in the directory after configuring the environment variables for the dependencies. $A_5$ describes how to execute the experiments.

**Artifact Execution**

DMRlib relies on the resource manager $A_2$ and the malleable application $A_4$, so both have to be running. The experiment workflow is detailed in $A_5$.

**Artifact Analysis (incl. Outputs)**

As in the previous sections, DMRlib relies on other subsystems as $A_5$ output shows.

### 2.2 Computational Artifact $A_2$

**Relation To Contributions**

$A_2$ is the malleable resource manager of $A_1$ based on Slurm. In this regard, we leveraged $A_2$ together with $A_1$ for designing the study case ($C_2$) and to validate MPI malleability in a supercomputer ($C_3$).

**Expected Results**

The malleable version of Slurm is capable of reallocates resources previously assigned to running jobs. We expect that $A_2$ determines the most appropriate number of resources for each job depending on the users specifications and the cluster status.

**Expected Reproduction Time (in Minutes)**

Likewise $A_1$, $A_2$ is active during the whole experiment. Its setup time is around 10 minutes.

**Artifact Setup (incl. Inputs)**

*Hardware.* Any computer.

*Software.*

(1) GCC v10.4
(2) OpenSSL v1.0.2j (https://github.com/openssl/openssl/tree/OpenSSL_1_0_2j)

*Datasets / Inputs.* Slurm inputs are the submitted jobs as described in $A_5$.

*Installation and Deployment.* The installation is done with:

```
./configure                                          \
    --prefix=$DMR_PATH/slurm-install                 \
    --sysconfdir=$DMR_PATH/slurm-confdir             \
    --without-pmix --with-ssl=$OPENSSL_PATH
make CFLAGS='-fcommon' CXXFLAGS='-fcommon'
make install
```

While the deployment is done during the experiments as shown in $A_5$.

**Artifact Execution**

$A_2$ relies on $A_1$ and the submitted jobs. The experiment workflow is detailed in $A_5$.

**Artifact Analysis (incl. Outputs)**

As in $A_1$, Slurm malleable relies on other subsystems as $A_5$ output shows.

## 2.3 Computational Artifact $A_3$
**Relation To Contributions**

$A_3$ is the workload submitter. It is responsible to extract users behavior and to adapt the information to the target cluster in order to be able to replay the workload ($C_1$) In this regard, we leveraged $A_3$ to validate MPI malleability in a supercomputer ($C_3$).

**Expected Results**

$A_3$ submits jobs to the queue handled by $A_2$ according to the input files.

**Expected Reproduction Time (in Minutes)**

$A_3$ is active during the whole experiment. The script is interpreted in execution time, so no compilation is required.

**Artifact Setup (incl. Inputs)**

*Hardware.* Any general-purpose computer that can run Python 3 interpreter.

*Software.* Python 3 interpreter.

*Datasets / Inputs.* $A_3$ expects users behaviors to be replayed. In the malleable experiments, it needs the behavior files for the *traditional users* and for the *generative user*.

*Installation and Deployment.* The Python scripts are executed during the experiments as $A_5$ describes.

**Artifact Execution**

$A_3$ submit jobs to $A_2$. The experiment workflow is detailed in $A_5$.

**Artifact Analysis (incl. Outputs)**

As in $A_3$ relies on other subsystems as $A_5$ output shows.

## 2.4 Computational Artifact $A_4$
**Relation To Contributions**

MPDATA is the scientific application executed by the PhD student. In this regard, it is essential in $C_2$.

**Expected Results**

MPDATA is leveraged to generate load for a given time depending on its configuration, so its results are irrelevant.

**Expected Reproduction Time (in Minutes)**

With the given configurations it can be completed between 36.5 and 1.45 hours, with 1 and 64 nodes, respectively.

**Artifact Setup (incl. Inputs)**

*Hardware.* Any computer.

*Software.* GCC compiler.

*Datasets / Inputs.* The input provided to MPDATA is a domain of $8,192 \times 1,024 \times 128$ cells that will run during $1,800$ steps.

*Installation and Deployment.* $A_4$ can be installed running make.

**Artifact Execution**

Since we only use one configuration of MPDATA, the generated binary can be launched as is.

**Artifact Analysis (incl. Outputs)**

The output is a message of "succesful execution" when finishing.

## 2.5 Computational Artifact $A_5$
**Relation To Contributions**

$A_5$ involves a series of scripts to validate MPI malleability in an actual systems ($C_3$) which coordinates the previous four artifacts.

**Expected Results**

$A_5$ provides all the necessary data to analyze the workload execution. Particularly, with $A_5$ can be created figures 6-15 and tables 1-2 and all the rational behind them.

**Expected Reproduction Time (in Minutes)**

Each experiments have run between 1.8 and 1.9 days.

**Artifact Setup (incl. Inputs)**

*Hardware.* Any HPC Cluster.

*Software.* Dependencies, scripts, and software in $A_1$, $A_2$, $A_3$, and $A_4$.

*Datasets / Inputs.* The experiments are launched with the inputs af the previous artifacts.

*Installation and Deployment.* Each experiment is launched with the `mnv_submission.sbatch` script configured with the desired strategy (see Section 4.4).

## Artifact Execution

A worklflow consists of the task chain: $T_1$, $T_2$, $T_3$, $T_4$, $T_5$, and $T_6$.

$T_1$ consists of a batch job that launches the nested malleable Slurm ($T_2$) and the user-based submitter ($T_3$). In turn, $T_3$ will submit replayed jobs ($T_5$) and malleable jobs ($T_4$) to the Slurm deployed in $T_2$. The malleable jobs submitted in $T_4$ are reconfigured in $T_6$ with DMRlib, which takes actions based on $T_2$ rational.

## Artifact Analysis (incl. Outputs)

$A_5$ generates all the necessary data to analyze the strategy evaluated in order to generete the paper results showed in sections 4 and 5.



\end{document}